\documentclass[compsoc,conference,a4paper,10pt,times]{IEEEtran}
\usepackage{cite}
\usepackage{amsmath,amssymb,amsfonts}
\usepackage{algorithmic}
\usepackage{graphicx}
\usepackage{textcomp}
\usepackage{bmpsize}
\usepackage{xcolor}

\usepackage[colorlinks=true,urlcolor=black,hidelinks]{hyperref}
\def\BibTeX{{\rm B\kern-.05em{\sc i\kern-.025em b}\kern-.08em
    T\kern-.1667em\lower.7ex\hbox{E}\kern-.125emX}}

\usepackage{subfigure}
\usepackage{fancyvrb}
\usepackage{tablefootnote}
\usepackage{blindtext,graphicx}
\usepackage[absolute]{textpos}
\begin{document}
% \bstctlcite{IEEEexample:BSTcontrol}

\begin{textblock}{17}(2,1.5)
\centering \noindent \large \emph{In Proceedings of the 7th IEEE EuroS\&P Workshop on Traffic Measurements for Cybersecurity (WTMC 2022)}
\end{textblock}

\title{Measuring and Clustering Network Attackers\\using Medium-Interaction Honeypots}

\author{\IEEEauthorblockN{Zain Shamsi\IEEEauthorrefmark{1}, Daniel Zhang\IEEEauthorrefmark{1}, Daehyun Kyoung\IEEEauthorrefmark{2}, and Alex Liu\IEEEauthorrefmark{1}\\}
	\IEEEauthorblockA{\IEEEauthorrefmark{1}Applied Research Laboratories, University of Texas at Austin, Texas, USA\\}
	\IEEEauthorblockA{\IEEEauthorrefmark{2}Dept. of Electrical and Computer Engineering,
		University of Texas at Austin, Texas, USA\\
		Email: \{zain, aliu\}@arlut.utexas.edu, daniel.zhang@live.com, danielkyoung2001@utexas.edu}
}

\maketitle
\thispagestyle{plain}
\pagestyle{plain}

\begin{abstract}
Network honeypots are often used by information security teams to measure the threat landscape in order to secure their networks. With the advancement of honeypot development, today's medium-interaction honeypots provide a way for security teams and researchers to deploy these active defense tools that require little maintenance on a variety of protocols. In this work, we deploy such honeypots on five different protocols on the public Internet and study the intent and sophistication of the attacks we observe. We then use the information gained to develop a clustering approach that identifies correlations in attacker behavior to discover IPs that are highly likely to be controlled by a single operator, illustrating the advantage of using these honeypots for data collection.
\end{abstract}


\hypertarget{introduction}{%
	\section{Introduction}\label{introduction}}

Computer and network security has become an extremely fast growing field
due to our world of connected devices. With individuals owning
multiple network-capable devices and organization networks consisting
of thousands, managing and securing them all is a tough undertaking for network
security teams. Along with large botnets continuously scanning network services for vulnerabilities, and attacks such as SolarWinds \cite{lazarovitz21} showing the effectiveness of supply chain attacks, security teams find themselves overwhelmed with attack vectors. Due to these events, many security professionals have adopted a view that a network breach is not an \emph{if} but a \emph{when}.

% This has made honeypots an important tool in the defense team arsenal,
% as they can provide important alerts from the pre-attack stage through post-intrusion. Network measurements using honeypots have generally used high-interaction honeypots (i.e., dedicated virtual machines) to capture as much data as possible, while low-interaction deployments (i.e., activity logging on open ports) have been used in practice due to ease of deployment and maintenance. However, in the last few years, we have seen a rise in the development of medium-interaction honeypots that can emulate a virtual environment for many different protocols as security professionals have realized the value of active defense.

Since most signature-based defense software rely on timely updates of their databases to counter known threats, and machine learning based systems can be evaded if their underlying models are not tuned correctly, attackers can find a window to sneak through the defense and establish their presence on the network. In these cases, honeypots buy time for the defense by slowing down the attackers who think they may have compromised real systems. Depending on the level of interaction provided by the honeypot or honeynet, this could not only increase the difficulty barrier for the attacker, but provide defenders a first look at the attacker's tactics and techniques.

Honeypots can be categorized based on the level of interaction they provide to the attacker.\textit{ Low-interaction honeypots} emulate an open service that an attacker can interact with, but provide limited emulation of the actual underlying software and operating system. The emulation is limited to a few protocol interactions and is used to log how the service is attacked using available protocol responses. \textit{Medium-interaction honeypots}, on the other hand, emulate a service and the system behind it with higher fidelity. This allows capturing logs of an attacker breaking the service as well as their activities once they believe they are on the system. The attacker is contained in a virtual environment, which they may eventually discover to be a honeypot as command responses are scripted and not every command may be implemented. Finally, \textit{high-interaction honeypots} are usually an entire virtual machine or a real machine deployed to work as a honeypot with a real open service. This level of deployment provides the highest fidelity, but is risky as the machine can be taken over completely by skilled adversaries and/or used to launch attacks from within the organization's IP space. 

% Network measurements using honeypots have generally used high-interaction honeypots to capture as much data as possible, while low-interaction deployments have been used practice for ease of deployment and maintenance. However, in the last few years, we have seen a rise in the development of medium-interaction honeypots for many different protocols as security professionals have realized the value of active defense.

% Another distinction made during deployment is the function of the
% honeypots. Though not mutually exclusive, honeypots deployed on the
% perimeter of the organization network are generally \emph{research}
% honeypots, with the goal being to gather intelligence about connection
% attempts made by scanners, inform about new vulnerabilities being
% exploited, and assist the security team in updating signatures to
% immunize the network against ongoing attacks. When deployed on the
% internal network, these \emph{production} honeypots imitate real
% services available inside the organization and serve as decoys to help
% defend against post-intrusion activity.

In this work, we investigate the use of several medium-interaction honeypots for characterization of network attacks. We deploy a set of honeypots spanning various protocols (i.e., SSH, HTTP, RDP, PSQL, PJL) on the public Internet and study the objectives of attackers that gain access to our honeypot virtual environments. We apply different measurement techniques that attempt to demystify attacker infrastructure and how the attacks originate. Armed with insights from our measurement, we then extract features we can use to cluster attacking IPs by building unique data representations for each feature. Finally, we combine our results from multiple clustering algorithms via a consensus clustering approach, ending up with a novel mechanism that can be used identify singular actors behind multiple attacking machines.

% Finally, we develop a novel technique to combine results from each feature via a consensus clustering approach that can be used identify singular actors behind multiple attacking machines.

\hypertarget{related-work}{%
\section{Related Work}\label{related-work}}

Network honeypots have been around since the early 2000s, initially started as projects by security professionals \cite{provos04,spitzner03}. Since then, researchers have recorded many measurements \cite{alata06, barron17, chong18, krupp16, krupp17, vetterl19, zeng14}, using honeypots to characterize attacks and activity captured.

Out of these previous works, the closest related to our work are the measurements done by \cite{barron17} and \cite{zeng14}. In \cite{barron17}, the authors deploy over a hundred SSH honeypots and attempt to quantify the behavior of attackers by varying the difficulty of entry and the content of the honeypot. They focus on measuring bot activity vs. human activity and provide a breakdown of the actions they observed under various scenarios. In \cite{zeng14}, the authors deploy a set of 18 high-interaction Windows virtual machines on Amazon Web Services (AWS) instances and report on how quickly they are attacked and taken over by malicious operators. They open all the ports on the machine and capture exploitation attempts on many common protocols as well as some unusual ones.

Since honeypots are a potent weapon in the network defender's arsenal, there has been a natural arms race between the attackers and defenders. Researchers working from the offensive perspective have offered up several ways to detect and avoid honeypots \cite{holz05, morishita19, vetterl18, vetterl19, zou06}, which has highlighted the weaknesses in default configurations and the work it takes to maintain a useful honeypot. Similarly, additional work on the defensive side has presented ways to attract attackers to trap them in a deceptive framework, such as leveraging social networks to post messages with exploitable information \cite{paradise17} and crafting responses to phishing emails \cite{koch13}.

There are several community led projects that seek to build better honeypots and provide data to security professionals to help investigate the latest attacks. The Honeynet Project \cite{spitzner03} has grown into an international organization that creates new honeypots, supports various research and runs an annual workshop. Other projects of note are Rapid7's Project Heisenberg \cite{heisenberg}, which maintains over 150 honeypots over the globe and provides some of their data available for download, and the STINGAR project \cite{biever21} maintained by Duke University which provides access to their crowdsourced dataset to university researchers.

\hypertarget{measurement-setup}{%
	\section{Measurement Setup}\label{measurement-setup}}

\begin{table}
	\renewcommand{\arraystretch}{1.3}
	\caption{Honeypot software}
	\label{tbl:software}
	\centering
	\begin{tabular}{llr}
		\hline
		Honeypot & Service & Interaction Level \\
		\hline\hline
		Cowrie \cite{cowrie} & SSH & medium \\ 
		miniprint \cite{miniprint} & printer & medium \\ 
		Sticky Elephant \cite{stickyelephant} & PSQL & medium \\
		Glastopf \cite{glastopf} & HTTP & low \\ 
		PyRDP \cite{pyrdp} & RDP & low \\
		\hline
	\end{tabular}
\end{table}

For our measurement, we identified and stood up several research honeypots on an AWS backbone. The honeypot software we identified for use are listed in Table \ref{tbl:software}. We chose these software because they were fairly popular and well maintained open source projects at the time of selection. Our honeypots were online from April 2019 to January 2021, with periodic downtimes for maintenance windows. All honeypots were hosted on their own AWS virtual machine with instance IPs and outbound connections disabled to prevent abuse. 

\hypertarget{cowrie}{%
\subsection{Cowrie}\label{cowrie}}

Out of all of these software, the Cowrie SSH Honeypot was the most well
maintained and configurable application. Due to this, and the popularity
of SSH as a remote service, we were able to conduct a more detailed set
of measurement experiments using Cowrie.

Cowrie allows a fairly extensive configuration of the virtual environment the user logs into. We set up this environment to look like an install of an Ubuntu 18.04 filesystem on an AWS instance, and added in various spreadsheet files containing bogus financial data to the user home directory. Along with this, we cloned various Github code repositories for parsing CSV files, to mimic a development machine. We also scripted popular linux system commands (e.g., \texttt{uname}, \texttt{ps}, \texttt{ifconfig}, \texttt{lscpu}, etc.) to respond as an AWS instance would.

Besides the filesystem, Cowrie also allows configuration of various authentication schemes for login (e.g., whitelisted passwords, random login). Throughout our measurement period, we changed our authentication schemes to perform different measurements, e.g., allowing only passwords known to be used by the Mirai botnet, allowing hosts to login after a random number of attempts, and allowing public key authentication. Finally, to expand further on this experiment, we set up three Cowrie honeypots in different AWS regions -- North Virginia, London, and Singapore to capture activity across continents.

% \begin{table}
% \caption{Subset of Mirai passwords}
% \label{tbl:mirai-passwords}
% \centering
% \begin{tabular}{lr}
% \hline
% Username & Password\\
% \hline\hline
% root & xc3511\\
% root & vizxv\\
% root & xmhdipc\\
% root & klv1234\\
% root & Zte521\\
% root & 7ujMko0vizxv\\
% root & 7ujMko0admin\\
% admin & 7ujMko0admin\\
% admin & meinsm\\
% \hline
% \end{tabular}
% \end{table}

% \begin{table}
% \caption{Common credential pairs removed}
% \label{tbl:common-creds}
% \centering
% \begin{tabular}{lr}
% \hline
% Username & Password\\
% \hline\hline
% root & root\\
% root & password\\
% root & 123456\\
% admin & admin\\
% admin & password\\
% admin & 123456\\
% \hline
% \end{tabular}
% \end{table}

\hypertarget{sticky-elephant-and-miniprint}{%
\subsection{Sticky Elephant and miniprint}\label{sticky-elephant-and-miniprint}}

Both Sticky Elephant (PSQL honeypot) and miniprint (print server honeypot) let us emulate an interactive interface with scripted responses where an attacker could attempt valid commands for each type of server. While they do not have any login authentication built in, they allow basic configuration changes to the default settings for reported versions of software and type of hardware device, which we ensured we changed to disable easy fingerprinting of the honeypots. Since these services were not as popular as Cowrie, and the attacks we saw did not attempt any intensive probing, we did not change our settings throughout the measurement period.

\hypertarget{glastopf-and-pyrdp}{%
\subsection{Glastopf and PyRDP}\label{glastopf-and-pyrdp}}

In order to follow good ethical research practices, we ran Glastopf and PyRDP as low-interaction honeypots to disable their misuse for further attacks. Glastopf was set up as HTTP honeypot to capture malicious web requests directed at our web server, and we disabled attacker interaction elements (e.g., user input forms and comment boxes) to prevent collateral damage (e.g., form hijacking, XSS attacks). Similarly, PyRDP was setup as a low-interaction proxy server to capture authentication events on the Windows remote desktop protocol, and we disabled access to a real Windows environment. For both honeypots, we again ensured that we modified the default settings, i.e., we changed the landing page for Glastopf, and set up a valid certificate presented to a RDP visitor.

\hypertarget{measurement-results}{%
	\section{Measurement Results}\label{measurement-results}}

Since information about specific vulnerabilities is readily available from better equipped crowdsourced honeypot projects, in our measurement we focus on insights that inform us about attacker motives and sophistication, as well as knowledge useful for building durable defenses.

\hypertarget{how-many-attacks-did-we-get}{%
\subsection{How many attacks did we get?}\label{how-many-attacks-did-we-get}}

\begin{table*}
    \caption{Cowrie honeypot activity across regions}
    \label{tbl:cowrie-overall-stats}
    \centering
    \begin{tabular}{|l|r|r||r|r|r|r|}
    \hline
         & Unique IPs & IPs with & Unique sessions & Sessions with & Sessions with & Sessions with \\ 
         & & successful login & & no login attempt & failed login & successful login \\
    \hline\hline
        %All regions & 73,142 & 4,829 (6.6\%) & 5,555,212 & 2,655,757 & 212,669 & 2,686,786 \\ 
        %N. Virginia & 46,028 & 1,899 (4.1\%) & 1,743,384 & 710,065 & 130,318 & 903,001 \\ 
        %London & 31,862 & 1,629 (5.1\%) & 2,026,146 & 787,328 & 55,403 & 1,183,415 \\ 
        %Singapore & 30,797 & 2,027 (6.5\%) & 1,785,658 & 1,128,063 & 57,225 & 600,370 \\ 
        All regions & 73,142 & 4,829 (6.6\%) & 5.55M & 2.66M (47.8\%) & 212.7K (3.8\%) & 2.69M (48.3\%) \\ 
        N. Virginia & 46,028 & 1,899 (4.1\%) & 1.74M & 710.0K (40.7\%) & 130.3K (7.5\%) & 903.0K (51.8\%) \\ 
        London & 31,862 & 1,629 (5.1\%) & 2.03M & 787.3K (38.9\%) & 55.4K (2.7\%) & 1.18M (58.4\%) \\ 
        Singapore & 30,797 & 2,027 (6.5\%) & 1.79M & 1.13M (63.2\%) & 57.2K (3.2\%) & 600.4K (33.6\%) \\ 
    \hline
    \end{tabular}
\end{table*}

Beginning with Cowrie, Table \ref{tbl:cowrie-overall-stats} shows the overall number of IPs and sessions we saw across all regions. Only a small fraction of IP addresses successfully logged in, and approximately half of the sessions were successful, which means that we observed a large number of return attempts from the same IP. This phenomenon is also observed in our session statistics, where for all attacker IPs the average sessions per IP was 75.95 and the median was 2. This large difference between the median and mean value tells us that most of the connection sessions were from a small subset of IP addresses that managed a successful login. Notably, 10 IP addresses with the most logins were responsible for over half of the total connections. However, due to NAT devices and IP churn over our measurement period, this could involve many more actual devices taking part in these attacks.

\begin{figure}
	\subfigure[Unique IPs across regions]
	{\includegraphics[width=1.612in]{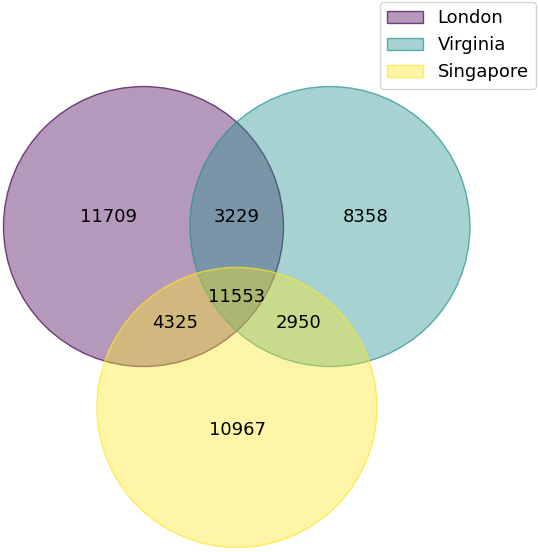}\label{Fig:ip-overlap}}
	\hfil
	\subfigure[Unique IPs across services]
	{\includegraphics[width=1.612in]{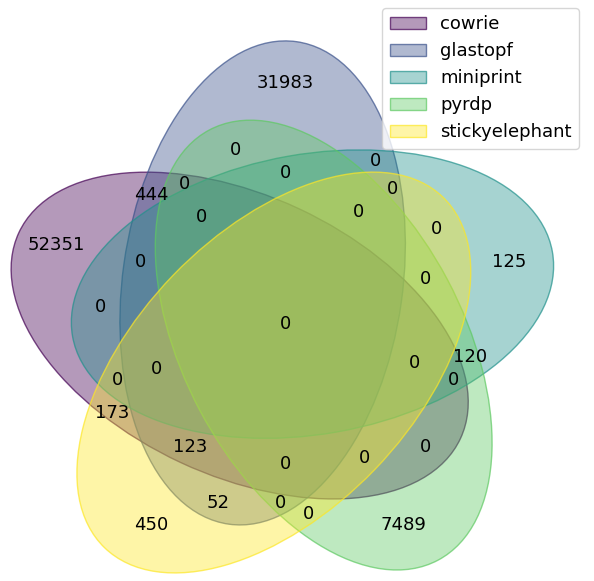}\label{Fig:service-overlap}}
	\caption{Activity overlap across honeypot services and Cowrie regions}
	\label{fig:overlaps}
\end{figure}

Next, we compare the attacks sustained by the individual Cowrie regions across the same period of activity. Figure \ref{Fig:ip-overlap} shows a Venn diagram of the overlap between the regions. We observed consistent numbers of overlapping attacker IPs across regions, but we also found each region to have a fairly large contingent of IPs unique to it. This was unexpected, as intuition suggests that Internet scanners generally distribute their targets uniformly, and assigning targets to specific regions would require too much coordination between scanning resources. However, this result points to scanners that may be localized to different parts of the world.

\hypertarget{tbl:honeypots-overall-stats}{}
\begin{table}
	\caption{Activity across other honeypots}
	\label{tbl:honeypots-overall-stats}
	\centering
	\begin{tabular}{|l|r|r|r|r|r|}
		\hline
		& Glastopf & PyRDP & StickyElephant & miniprint \\ 
		\hline\hline
		% Unique IPs & 73,142 & 34,937 & 13,431 & 1,375 & 405 \\ 
		% Total connections & 5,555,212 & 405,532 & 1,018,982 & 13,501 & 1,845 \\
		Unique IPs & 34.9K & 13.4K & 1.3K & 405 \\ 
		Connections & 405.5K & 1.02M & 13.5K & 1.8K \\ 
		\hline
	\end{tabular}
\end{table}

Table \ref{tbl:honeypots-overall-stats} shows the number of unique IPs
and total connections we observed on all the other honeypot services. These services saw a much lower amount of traffic, however they exhibited a similar trend: a majority of connecting IPs were only seen once, and a few IPs were responsible for the majority of repeated connections. We also checked the number of unique services that each attacking IP attempted to access, shown in Figure \ref{Fig:service-overlap}. While there were a few IPs seen targeting two or three services, this was a rare event and most attackers overwhelmingly focused on a single service.

% \begin{figure}
% \centerline{\subfigure[Case I]{\includegraphics[widt
% h=2.5in]{subfigcase1}
% \label{fig_first_case}}
% \hfil
% \subfigure[Case II]{\includegraphics[width=2.5in]{su
% bfigcase2}
% \label{fig_second_case}}}
% \caption{Activity overlap across regions and services}
% \label{fig:overlaps}
% \end{figure}

\hypertarget{how-do-the-attacks-originate}{%
\subsection{How do the attacks originate?}\label{how-do-the-attacks-originate}}

Since IPs from any part of the world can fairly easily be obtained, we surmise that any actor engaging in malicious scanning would leverage rented IP address space. By studying the ASes of the attacker IPs on our Cowrie honeypots, we found no surprises in that they were mostly composed of large cloud providers such as DigitalOcean, OVH, Tencent Cloud, and Azure, along with IPs from Chinese ASes that produced most of scanning activity.

% We also take a look to determine if attackers use the TOR anonymity
% network to scan anonymously. We only found 86 TOR IPs on our London
% Cowrie honeypot, 111 on Singapore, and 84 on N. Virginia, which are all
% less than 0.3\% of the IPs observed. This makes sense, as the slow
% speeds of the TOR network severely throttles attackers from quickly
% scanning the Internet.

\begin{figure}
	\subfigure[N. Virginia region only]
	{\includegraphics[width=1.61in]{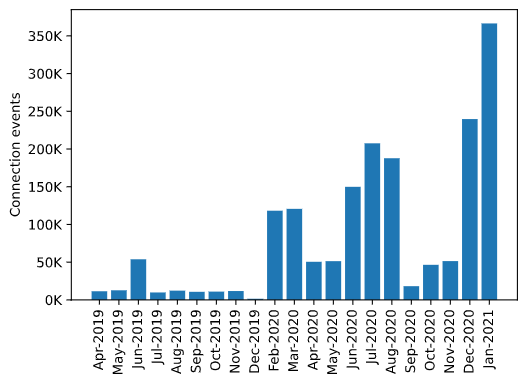}}
	\hfil 
	\subfigure[Across all regions]
	{\includegraphics[width=1.61in]{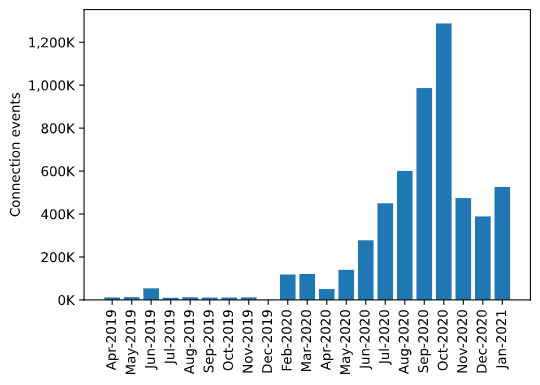}}
	\caption{Increase in connection events observed on Cowrie honeypots}
	\label{Fig:attacks-over-time}
\end{figure}

A more interesting statistic surfaces when examining the attacks over time. We noticed that there was a big increase in attacks when comparing activity across 2019-2021. Figure
\ref{Fig:attacks-over-time} shows a significant increase in overall
connection events heading into 2020. While this is difficult to attribute to
a specific attack or vulnerability, we can point to increased global
network activity in general due to remote work during the COVID-19
pandemic, with more incentive for malicious actors to find machines with open ports on the Internet, as well as more infected machines to perform the scans.

% \begin{figure}
% \hypertarget{Fig:attacks-over-time}{%
% \centering
% \includegraphics[width=2.5in]{images/nvirg_connections.png}
% \caption{Connection events on Cowrie honeypot in N.Virginia over time}
% \label{Fig:attacks-over-time}
% }
% \end{figure}

In November 2020, we ran a small experiment to restart our honeypots periodically to see if new IP addresses assigned to the honeypots broke the prevalent trends in attack patterns. We found that the activity quickly rose to previous levels within a few days of the restart, but this could explain some of the decline seen in plot. Additionally, we observed a small set of IP addresses that were only seen once after each restart, which could point to a certain level of sophistication in these attackers that had some way to determine our machines to be honeypots and avoid them. We found no session activity from these IPs in our logs, nor were our honeypots marked with a high HoneyScore \cite{honeyscore}, however it is possible for there to be some markers in the protocol handshake stages which we did not observe as we did not have full packet captures.

In the final phase of our Cowrie measurement, we attempted to measure if attackers use leaked private keys scraped from paste sites. During the one month period of this measurement, we found no attackers using our leaked keys, though we discovered other keys being attempted. This points to the possibility of attackers injecting public keys into machines they gain access to. While we did not make our \texttt{\textasciitilde{}/.ssh} folder accessible on our honeypots, this would make for an interesting future measurement.

\hypertarget{what-do-attackers-do-once-connected}{%
\subsection{What do attackers do once connected?}\label{what-do-attackers-do-once-connected}}

The activity capture from the honeypots is likely the most interesting piece of the measurement, and we use this information to extract the attackers' intentions for executing their attacks.

\subsubsection{Cowrie}
Starting with Cowrie, we found that the attacker activity seen across all three regions was similar in nature, hence we grouped all activity together for this section. Beginning with the first short phase of our experiment where only known Mirai credentials were allowed to login, we found 31 out of 47 sessions with commands in this phase used identical sequences of known Mirai commands (listed in Appendix \ref{session-activity-appendix}). The remaining 16 sessions used slight variations of the known Mirai commands, revealing the presence of Mirai variants attempting to create shadow botnets by re-using the Mirai credentials.

After this first phase, we allowed attackers to login after a random number of attempts (i.e., 6-10) for the majority of our measurement period. We noticed that a large number of sessions that successfully logged into our Cowrie honeypots immediately made an outbound connection, while a smaller fraction used shell commands to manipulate the virtual filesystem. Since our honeypots were open machines on a random AWS instance, we conjecture that most attackers would constitute of automated bots seeking to recruit additional zombies for their botnet, run DDoS or reflection attacks, or set up web/mail proxies and tunnels for personal or commercial use. Previous measurements have also shown companies posing as legitimate proxy providers using hacked machines as part of their service \cite{mi19}. 

This explanation is validated by observing the port numbers and requested domains of the outbound requests, with the top 10 most popular shown in Table \ref{tbl:outbound-requests-numrequests}. Out of all requests, 85\% were to web services via standard ports 80 and 443, which included homepages of Google and Yandex presumably as connectivity checks, requesting various APIs for online services (e.g., Evernote, Amazon, Walmart) most likely to crawl them or send spam emails, or accessing media on YouTube and VK to increase viewer counts. While not shown in this table, one of the frequent URLs requested (api.bablosoft.com) gives a hint as to the intentions of some of the attackers, as this website sells a bot service to increase viewership on streaming platforms (e.g., YouTube, VK, Twitch). In the table, we also observe a small number of IPs creating a large number of requests for content hosted on a domain we anonymize as Organization1, as well as destinations unmapped to any DNS record which we discover belong to the same group. We study this organization in detail later in Section \ref{case-study-organization1}. Out of the remaining types of requests, 13\% of outbound requests were to addresses hosting mail services via standard ports 25, 110, 465, 587, and 993. The remaining 2\% used unusual ports to connect to IP addresses that hosted malware content.

% To better understand the distribution of the domains requested by outbound
% requests, observe , which
% shows some of the top outbound requests observed with the top-level
% domain extracted from the request and sorted by the number of requests.
% We can observe a variety of intentions from our attackers, such as
% accessing the homepages of Yandex (ya.ru) and Google presumably as a
% connectivity check, and accessing media on Youtube and VK. Note that we
% also see a small number of IPs creating a large number of requests for
% private content hosted on IP addresses without a domain, which are
% linked to the zennolab domain. We study this domain in detail later in
% Section \ref{case-study-zennolab}. 

\hypertarget{tbl:outbound-requests-numrequests}{}
\begin{table}
	\centering
	\caption{Top outbound requests}
	\label{tbl:outbound-requests-numrequests}
	\begin{tabular}{|l|r|r|r|}
		\hline
		Request domain & Port & Number of requests & Unique Ips \\ 
		\hline\hline
		ya & 80 & 1,952,086 & 125 \\ 
		Organization1 & 80 & 1,142,030 & 44 \\ 
		youtube & 443 & 471,730 & 126 \\ 
		google & 443 & 309,723 & 1,867 \\ 
		185.143.172.66 & 80 & 149,477 & 7 \\ 
		vk & 443 & 145,019 & 56 \\ 
		evernote & 443 & 137,206 & 17 \\ 
		amazon & 443 & 98,140 & 391 \\ 
		walmart & 443 & 91,116 & 39 \\ 
		163.172.20.152 & 80 & 89,055 & 9 \\
		\hline
	\end{tabular}
\end{table}

Next, we examine the commands executed on the SSH honeypots. Across
all phases, a total of 4,019 IP addresses used a shell command across
49,495 sessions. The majority of the sessions consisted of a single
command, with only 5,807 sessions from 3,225 attackers having multiple
commands. Focusing our analysis on these 5.8K sessions, we found
144 different command sequences, with 46 shared by multiple attackers. A
large number of attacker IP addresses (1,367 of them) had a shared command
sequence attempting to identify Microtik routers \cite{emily21}, which was a prominent attack during our period of measurement. Other command sequences (listed in Appendix \ref{session-activity-appendix}) show attempts to move into the \texttt{/tmp} directory to download and execute files, wipe all history and logs from the machine before adding a new user, and just generally output information about the system.

% Table \ref{tbl:command-popularity} shows the popularity of
% commands executed on the honeypots. Since most of our attackers were
% bots running automated scripts, it follows that we see almost all
% activity using the \texttt{echo} command to redirect input and create
% files. The other popular commands are for discovery of the system
% (\texttt{uname}, \texttt{cat}, \texttt{ps}) and for covering tracks
% (\texttt{unset}, \texttt{history}).

% \hypertarget{tbl:command-popularity}{}
% \begin{table}
%     \caption{Command popularity on honeypots}
%     \label{tbl:command-popularity}
%     \centering
%     \begin{tabular}{|l|r|}
%     \hline
%         Command & Sessions with command \\ 
%     \hline\hline
%         echo & 43,885 \\ 
%         uname & 7,239 \\ 
%         sudo & 3,285 \\ 
%         cat & 2,629 \\ 
%         unset & 2,004 \\ 
%         history & 2,001 \\ 
%         export & 1,996 \\ 
%         ls & 1,769 \\ 
%         ps & 1,746 \\ 
%         cd & 1,716 \\ 
%     \hline
%     \end{tabular}
% \end{table}

Finally, we study sessions that had human activity by using the Cowrie TTY logs, which record the timestamp of every key input for each interactive session. To flag sessions as human, we employed the three techniques from \cite{barron17}, which are (1) the presence of a backspace or delete character, (2) the maximum time delta between keystrokes is an outlier across all sessions, or (3) the maximum time delta is greater than 0.1 seconds. This yielded 46 sessions flagged as human activity, which is a very low percentage out of 2.6M successful logins (0.0017\%), and 49K sessions with shell commands (0.09\%). 

Our human classified sessions used 6.39 commands on average with the most being 22 commands. They predominantly used commands for discovery (e.g., \texttt{ifconfig}, \texttt{uname},
\texttt{arp}), navigating the filesystem (e.g., \texttt{ls}), and file
download (e.g., \texttt{curl}, \texttt{wget}). Examining
the sessions manually, we found users interested in our planted files,
browsing to our fake financial data folder and attempting to open the
data files, while others were interested in installing scanning tools
such as Nmap and Masscan to run Internet scans. A few users
attempted to add additional users and change the account password to set
up persistence, while a couple simply attempted to \texttt{rm\ -rf} the
entire machine. Since Cowrie provides a virtual environment for each
attacker with scripted responses, we noticed the humans becoming
frustrated at commands not causing any changes on the system and exiting the session prematurely. This activity capture shows that Cowrie does well in entrapping human attackers, and the virtual environment of a medium-interaction honeypot provides protection against attackers seeking to change system files, add users and software, or just wipe the machine.

\subsubsection{Sticky Elephant} 
From the 1,375 IP addresses we observed on StickyElephant, 485 requested a pre-login handshake, 64 attempted login with password, and only 12 attempted SQL queries. Out of these, 3 IP addresses made up most of the query attempts with almost 2000 queries that attempted to brute-force passwords, use PSQL extensions to run system commands, and lookup the database schema to search for unprotected tables. Our third attacker was perhaps the most interesting, as not only did they attempt to output the database structure, but made several queries to look for hidden schemas, grab table names and various
statistics about table sizes, and followed up by dropping all tables
and inserting a ransomware message holding the data hostage (listed in Appendix \ref{psql-activity-appendix}). However, closer inspection of the log data revealed that the attacker did not run any commands to exfiltrate data from the database prior to deleting the tables and delivering the fraudulent ransom request. 

% \begin{figure}
% \hypertarget{Fig:ransomware}{%
% \centering
% \includegraphics[width=3.5in]{images/psql-ransomware.png}
% \caption{Ransomware message left on PSQL honeypot}
% \label{Fig:ransomware}
% }
% \end{figure}

\subsubsection{miniprint}

% \begin{table*}
%     \caption{Activity breakdown on miniprint honeypot}
%     \label{tbl:miniprint-activity}
%     \centering
%     \begin{tabular}{|l|r|r|r|r|r|r|r|r|}
%     \hline
%          & Raw text print & PJL commands & ID/Status & Filesystem query & Echo request & Console ready & Turn off & PS print \\ 
%      \hline\hline
%         Observed & 1676 & 615 & 430 & 74 & 51 & 37 & 23 & 12 \\ 
%         UniqueIPs & 181 & 100 & 89 & 2 & 10 & 14 & 16 & 3 \\ 
%     \hline
%     \end{tabular}
% \end{table*}

\hypertarget{tbl:miniprint-activity}{}
\begin{table}
    \caption{Activity breakdown on miniprint honeypot}
    \label{tbl:miniprint-activity}
    \centering
    \begin{tabular}{|l|r|r|}
    \hline
    Command          & Observed & Unique IPs \\
    \hline\hline
    Raw text print   & 1676     & 181       \\
    PJL commands     & 615      & 100       \\
    ID / Status      & 430      & 89        \\
    Filesystem query & 74       & 2         \\
    Echo request     & 51       & 10        \\
    Console ready    & 37       & 14        \\
    Turn off request & 23       & 16        \\
    Postscript print & 12       & 3         \\
    \hline
    \end{tabular}
\end{table}

Our printer honeypot saw a considerably lower number of attackers and connections compared to other honeypots, but still saw a variety of printer commands being attempted. Table \ref{tbl:miniprint-activity} shows the breakdown of this activity. As expected for a honeypot simulating a network connected printer device, the most common action observed was raw text print, with 1,676 print requests from 181 IP addresses, followed by attempts to fingerprint the device and turn off the printer. There were only 19 IP addresses that made both print and PJL command actions, showing that the majority of attackers focused on using either PJL commands or attempting to print. Of the print attempts, we noticed activity ranging from general vandalism to advertisements (e.g., for network security services), and political hacktivism such as attempts to print messages about injustice in HongKong and the Black Lives Matter movement (examples in Appendix \ref{miniprint-activity-appendix}). 

\subsubsection{Glastopf} 
Our Glastopf web honeypot saw a total of 34,937 unique IP addresses who sent 405,532 requests. As is par for the course in network measurements, our distribution followed a heavy tail, with the large majority of IPs sending a small number of web requests, with 22,081 IP addresses having only one request and 93\% of IPs having ten requests or less. Most requests we received were GET requests, with only 4\% of overall requests being one of POST, HEAD, PUT, or OPTIONS. 
%32,473 = 93%

To determine what our visitors were trying to achieve, we investigate the common GET requests seen in the Glastopf logs, shown in Table \ref{tbl:glastopf-activity} with duplicates removed. We see a variety of objectives for the attackers, from just browsing the homepage, looking for CGI scripts and restricted pages using robots.txt, attempts at various vulnerabilities, and command execution. We discovered 2 notable IPs that repeatedly attempted to access AWS meta-data, mimicking the Capital One hack in 2019 \cite{neto20}. This attack is only applicable on AWS machines, which means that these particular attackers were aware of and specifically targeting the AWS IP space.

\hypertarget{tbl:glastopf-activity}{}
\begin{table}
	\caption{Popular requests seen on glastopf honeypot}
	\label{tbl:glastopf-activity}
	\centering
	\scriptsize
	\begin{tabular}{|l|r|r|r|}
		\hline
		Request URI & Requests & IPs & Vulnerability \\ 
		\hline\hline
		/ & 9669 & 2769 & None (crawl) \\ 
		/dynamic/instance-identity/document & 1154 & 2 & AWS meta-data\\ 
		/manage/cgi/ & 659 & 531 & CGI scripts \\ 
		/robots.txt & 620 & 235 & Restricted files \\ 
		http://169.254.169.254/ & 577 & 2 & AWS meta-data \\ 
		/pma2012/ & 556 & 16 & PHPMyAdmin \\ 
		/phpmyadmin/4.2/installing/ & 551 & 12 & PHPMyAdmin \\ 
		/.env & 477 & 301 & Command exec \\ 
		/FxCodeShell.jsp/AT-generate.cgi & 364 & 11 & Tomcat \\ 
		/TP/public/index.php & 301 & 272 & ThinkPHP \\ 
		\hline
	\end{tabular}
\end{table}

\subsubsection{PyRDP}

\hypertarget{tbl:pyrdp-channels}{}
\begin{table}
	\caption{Virtual channel extensions bound by RDP attackers (rdpdr=filesystem, cliprdr=clipboard, rdpsnd/snddbg=sound output, drdynvc=dynamic channel, MS\_T120=BlueKeep)}
	\label{tbl:pyrdp-channels}
	\centering
	\begin{tabular}{|l|r|r|}
		\hline
		Virtual Channel extension & Sessions & Unique IPs \\ 
		\hline\hline
		MS\_T120 & 435 & 90 \\ \hline
		rdpdr, drdynvc, cliprdr, rdpsnd & 54 & 49 \\ \hline
		rdpdr, cliprdr, rdpsnd, snddbg & 113 & 61 \\ \hline
		rdpdr, MS\_T120, cliprdr, rdpsnd, snddbg & 15 & 11 \\ 
		\hline
	\end{tabular}
\end{table}

Our RDP honeypot had the second greatest number of connections and unique IP addresses that connected to it following the Cowrie honeypots, with over 1M connections from 13,431 IP addresses. However, only 46 connections, originating from 40 attacker IP addresses, resulted in the client attempting to log in with credentials. Out of these, 35 attacker IP addresses attempted blank credentials, while the other 5 attackers attempted login using common administrator credentials.

The RDP protocol uses multiple virtual channels for communication between the server and client. Several extension channels have been added for functionality as the protocol has evolved (e.g., sound output, clipboard access). Table \ref{tbl:pyrdp-channels} shows the distribution over 617 connection sessions that bound these extension channels. The PyRDP logs identified one type of attack, BlueKeep, which works by causing the server to bind the MS T120 virtual channel - an internal channel that no client should connect to.

We saw 450 attempts at BlueKeep originating from 101 IP addresses, with most simply attempting to bind only to the internal channel, but 11 IPs including other known channels. Since these extended virtual channels are optional, observing which virtual channels are bound by the attackers can give us a fingerprint for grouping attackers using similar methods of attack, and this has been explored by security researchers to identify attack tooling \cite{ka19}. For example, the 11 IPs in the last row of the table match the Metasploit scanner for BlueKeep. Unfortunately, besides this match we found no openly available signature database of RDP fingerprints that we could use to identify our other attackers, and we leave this as an avenue for future work.

\hypertarget{clustering-similar-attackers}{%
	\section{Clustering Network Attackers}\label{clustering-similar-attackers}}

In this section, we use the knowledge gained from our measurement to extract various features that can be used to group attacking IPs on our Cowrie honeypots. We develop unique representations of the data for each feature, and apply standard clustering algorithms to find an initial set of groupings. Next, we draw from previous work in consensus clustering to develop a method to combine our results while filtering out inaccuracies using a greedy optimization approach. Since our measurement showed each region to have a large set of unique IPs observed, we keep our clustering separate for the three Cowrie regions.

% In this section, we use the knowledge we gained from our measurement to study correlations in attacker activity on our Cowrie honeypots and develop a novel approach to discover IPs that behave similarly. Specifically, we investigate various features that can be used from our honeypot logs and how they can be applied to find groupings in the data. Next, we draw from previous work in consensus clustering to develop methods to combine their results by filtering out inaccuracies using a greedy optimization approach. Since our measurement showed different sets of unique IPs observed on the Cowrie honeypots, we keep our clustering separate for the three regions.

% \subsection{Independent clustering over features}

% To begin, we build different sets of clusters using various features collected from the Cowrie logs. We treat each feature separately as each works over a different set of attackers and requires a tailored approach.

\subsection{Independent Clustering over Features}

We begin by treating each feature separately, as each works over a different set of attackers and requires a tailored approach.

\hypertarget{heuristic-grouping}{%
	\subsubsection{Heuristic Grouping}\label{heuristic-grouping}}

We start with employing a simple heuristic approach as a baseline for
grouping attackers. This method relies on our domain knowledge as well
as our insights from our measurement above. Since we found most of our
attackers to be automated bots, we base our heuristics on their scripted
behavior. Defining a graph where nodes are attacking IPs, we draw an
edge between two nodes if it matches any of these three criteria:

\begin{enumerate}
\item
  Both nodes have exactly the same set of credentials attempted, given both have
  each attempted at least 5 different credentials
\item
  Both nodes have exactly the same sequence of commands executed in a session,
  given each of the sessions contain at least 5 commands
\item
  Both nodes create the same outbound request upon login, where the
  request is to an unusual port. As we saw in our measurement above, we
  had about 2\% of IPs attempt a request to a port that was not web or
  email.
\end{enumerate}

The clusters are then formed by taking all connected components from the
graph. This results in us finding 21 clusters on our London data, 22 in
Singapore, and 36 in N. Virginia. Given the strict criteria for matching, we gain a set of clusters in which we have a high confidence of accuracy. Additional information can be gained by relaxing the criteria, which we do in the following sections.

\hypertarget{outbound-requests}{%
\subsubsection{Outbound Requests}\label{outbound-requests}}

As we found during our measurement, intruders were most likely to create
outbound requests to remote servers as part of a post-compromise
routine. Because our Cowrie instances disabled creation of outbound connections, we were
unable to observe complete interactions between attackers and their requested servers, but we were able to capture the connection attempt. Observing different attacker IPs attempting to connect to the same external host to reach out to an API or download malware gives us a hint to their purpose, and can be used as a means for grouping them together. In Table \ref{tbl:outbound-requests-numrequests}, we noticed that most attackers attempted connections to various web and email ports. We used the 2\% of IPs that connected to unusual ports instead as part of our heuristic grouping, but here we use the entire set of outbound request information to create clusters.

% \begin{figure}
% \hypertarget{Fig:outbound-singapore}{%
% \centering
% \includegraphics[width=3.5in]{images/outbound-request-clusters.png}
% \caption{Graph built from outbound requests observed in the Singapore region.}
% \label{Fig:outbound-singapore}
% }
% \end{figure}

\begin{figure}
	\subfigure[Singapore outbound requests]
	{\includegraphics[width=1.6125in]{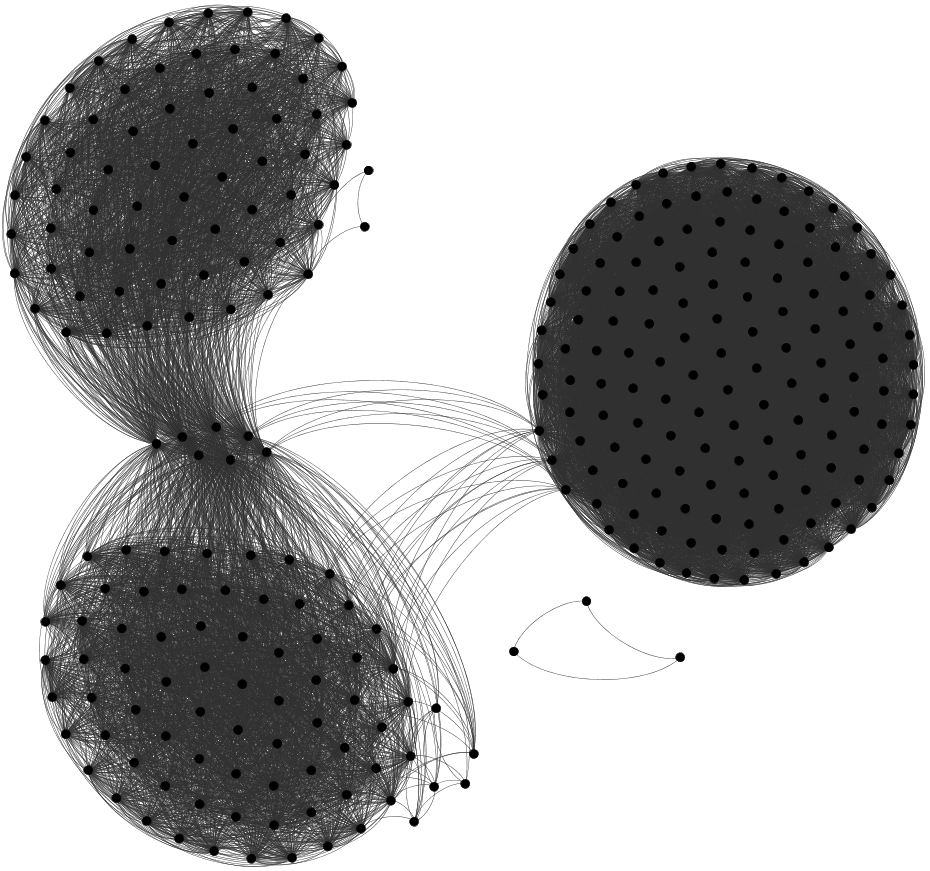}\label{Fig:outbound-singapore}}
	\hfil 
	\subfigure[London shared credentials]
	{\includegraphics[width=1.6125in]{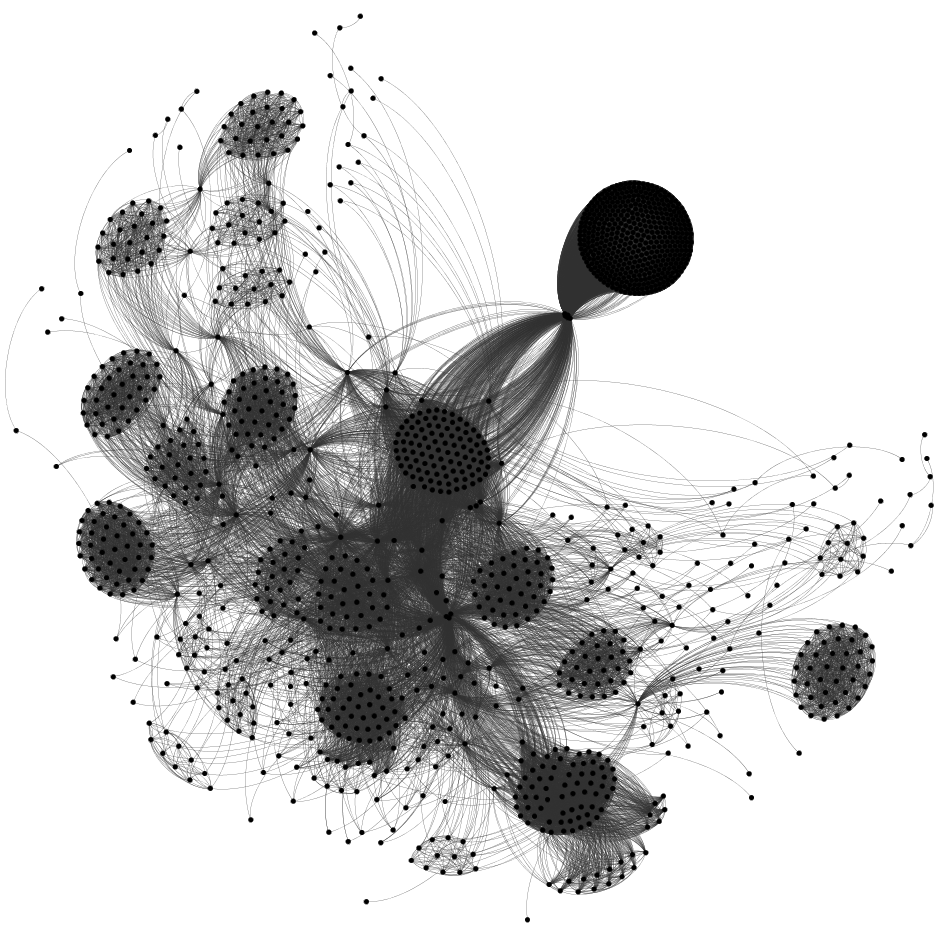}\label{Fig:sharedcreds}}
% 	\subfigure[Timeseries of connection events]
% 	{\includegraphics[width=1.61in]{images/probe-interval-cluster.png}\label{Fig:interval-cluster}}
	\caption{Example visualizations from independent feature clustering}
	\label{Fig:example-graphs}
\end{figure}

% \begin{figure}
% \hypertarget{Fig:sharedcreds}{%
% \centering
% \includegraphics[width=3.5in]{images/shared-credential-communities.png}
% \caption{Shared credential graph built from London data}
% \label{Fig:sharedcreds}
% }
% \end{figure}

We constructed a graph where unique IPs represent nodes, and an edge is formed between two nodes if they have requested at least one domain in common. We next considered weighting each edge by the number of requested domains in common and the time period during which they were requested. However, we quickly found that these weights did not make much of a difference when it came to the clustering results, as there was very little overlap in the domains requested, and all three regions had well defined clusters that surfaced quickly. An example of the graph built for the Singapore region is shown in Figure \ref{Fig:outbound-singapore}. We obtained our clusters by using spectral clustering on the graph adjacency matrix, choosing \(k\) based on the eigengap metric \cite{vonlux07}, and visually inspecting for correctness. We end up with 5 clusters each for the London and N. Virginia regions, and 4 for Singapore.

\hypertarget{probe-intervals}{%
\subsubsection{Probe Intervals}\label{probe-intervals}}

Internet scanning activity is often pre-configured with sweeps of the IP space occurring on regular intervals. While determining coordination between distributed scanners is difficult and still an open research problem \cite{bhuyan11}, here we follow the assumption that malicious distributed scanning setups (e.g., botnets) are often made up of automated zombie hosts (e.g., hacked devices and cloud instances) that can exhibit too much churn in their uptime \cite{herwig19, stone09} for a controller to continuously optimize a coordinated scanning approach. Thus, the most practical approach is to simply assign each host a similar list of scan targets and ask them to scan as much as possible. This can result in our honeypots observing multiple IPs completing numerous sweeps following a similar scanning schedule.

Thus, we attempt to use intervals between probes seen from individual IPs as a way to cluster those IPs that may be configured with a similar scan schedule. We parsed our honeypot logs for all attempted connections, disregarded whether the connection attempt succeeded or
failed, and logged all timestamps generated by an IP, creating a list of intervals. We then
projected each interval list into a timeseries spanning the entire length of the period the IP was observed in our measurement. We further subdivided each timeseries into two week windows to ensure we capture the temporal locality of the scanners, as well as for more manageable computation.

To cluster our set of timeseries, we used Dynamic Time Warping (DTW) as a similarity metric. DTW has the benefit of allowing shifted timeseries to still be matched, which was important in our case as our scanning IPs are likely not synchronized in their scanning patterns. Once we obtained a distance between all pairs of interval lists within each window, we clustered them using the OPTICS algorithm \cite{ankerst99}. Finally, we combined the result of clustering over all windows of the measurement period by deriving a final distance between two IPs as the fraction of how many times the pair was clustered when they both appeared within the same window. 

\begin{figure}
\hypertarget{Fig:interval-cluster}{%
\centering
\includegraphics[width=2.4in]{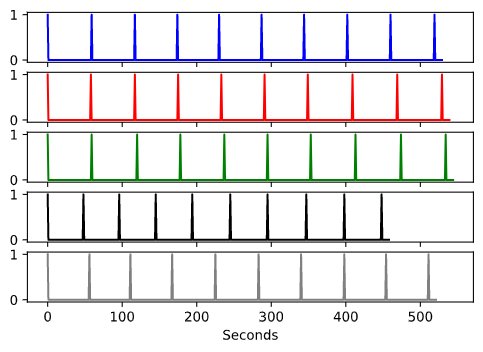}
\caption{Timeseries of probe events from one of our London clusters.}
\label{Fig:interval-cluster}
}
\end{figure}

Because of the large variance in probe intervals across our data, we found that many attackers with several seconds of distance in their sequence were eventually clustered together. Figure \ref{Fig:interval-cluster} shows five such attacker connection event sequences from one of our clusters in the London data, where the spikes indicate a probe. Observe that the intervals can be shifted, may differ by several seconds, and the sequence may terminate earlier or later. We end up with 21 clusters from the London data, 27 in the N. Virginia, and 29 from the Singapore region.

\hypertarget{session-activity}{%
\subsubsection{Session activity}\label{session-activity}}

Access to the captured command history of the attacker is one of the primary advantages of using medium-interaction honeypots for such a measurement. This points to the attacker's purpose, and provides a compelling piece of information for grouping attackers. Since we have already captured IPs that run the exact same script as part of our heuristic grouping, in this section we take a look beyond the exact commands executed, and attempt to infer the intentions of the attackers and cluster them based on this metadata.

We started by gathering all sessions in which there was command input and then reduced each command to the binaries executed in the command by dropping arguments, the \texttt{sudo} keyword, and breaking up multiple statements. This left us with a mapping describing each session by a sequence of executed binaries. We then infer the \textit{intent} of the activity from the sequence of binaries using GTFOBins \cite{gtfo}, which is a curated list of linux commands and how they can be abused on systems. More specifically, this list maps a set of common linux commands to a set of possibly malicious capabilities (e.g., file upload, sudo access, reverse shell). By clustering on this metaset of capabilities instead of the commands themselves, we can uncover similarities between attackers with the same objective, but different command executions.

To cluster on the capability set, we replaced each command in our data with an \textit{embedding}, represented as a binary vector that marks the capabilities of each command, and padded each session to the same length by adding zero vectors. To determine similarity between sessions, we calculated the distance between the embeddings in sequence and then averaged the result. This was done to maintain the sequential information of the executed commands. Once we had a pairwise distance for each of our sessions, we use the OPTICS algorithm to find clusters of sessions and mapped them back to the originating IP. We end up with 10 clusters from the London region, 11 out of Singapore, and 25 out of N. Virginia.

Our method correctly clusters sessions in which the IPs had the same sequence of binaries, which is the straightforward case. We also end up with a few clusters in which the sequence is not exactly the same (e.g., a few commands are skipped), and a few small clusters where the match was using the capability set. Table \ref{tbl:session-clusters} shows truncated examples from two such clusters where cluster $A$ had three different attackers reading files, querying network configuration, and then system information with slight variations. In cluster $B$, the attackers performed operations on the directory, followed by system discovery and a file download.

\hypertarget{tbl:session-clusters}{}
\begin{table}
    \caption{Example commands from sessions clustered in the N. Virginia data}
    \label{tbl:session-clusters}
    \centering
    \begin{tabular}{|l|l|}
    \hline
        Cluster & Commands \\ 
    \hline\hline
        $A$ & \texttt{cat, echo, ifconfig, ip, ls, ps, uname} \\ 
        $A$ & \texttt{cat, ifconfig, ip, uname} \\
        $A$ & \texttt{cat, ifconfig, ip, ps, uname} \\ 
    \hline
        $B$ & \texttt{ls, nproc, wget} \\
        $B$ & \texttt{cd, chmod, rm, uname, wget} \\
        $B$ & \texttt{ls, lscpu, passwd, wget} \\ 
    \hline
    \end{tabular}
\end{table}

\hypertarget{credential-lists}{%
\subsubsection{Credential lists}\label{credential-lists}}

SSH brute force attacks have been common place on Internet-accessible SSH servers for several years \cite{javed13}. For any such attack to be feasible, the attacker usually specifies a list of credentials that the scanner can attempt. This list is generally copied to all the scanners, due to the difficulties in achieving coordination that we discussed in Section \ref{probe-intervals}. Thus, grouping our attackers based on credentials attempted is an effective way to discover clusters within our gathered set of IPs. 

We considered IPs with exact sets of matching credentials and included them in our heuristic grouping above. Here, we relax the criteria and match credentials attempted by all attackers to each other. We gathered all login attempts that were observed in the logs, and built a set of attempted credentials mapped to each attacking IP. The main complication in this approach is that the sets may have arbitrary overlap due to uncoordinated scanning done by several hosts belonging to the same attacking group, as well as overlap in credentials attempted by unrelated attackers.  Thus, we use the jaccard distance between two sets as our metric and then compute clusters using the OPTICS algorithm.

The one drawback of using the jaccard distance is that small sets of
common credentials (e.g., admin:password) get clustered together, and we have no additional information in the credential approach to discern if these attackers should really be
grouped. However, if the sets of attempted credentials are unique enough, jaccard works well to find meaningful clusters. We end up with a large cluster for each region
that captures the common credentials, and several other smaller clusters
where the IPs had significant overlap in the attempted credential set. We
find 13 clusters in London, 19 in Singapore, and 21 in N. Virginia. Note that we could drop the large common credential cluster at this point to remove some inaccuracies, but we leave the error correction for each of the independent clustering methods to our combining step.

\hypertarget{credential-sharing}{%
\subsubsection{Credential Sharing}\label{credential-sharing}}

We observed in our measurement that we had a large number of sessions where attackers disconnected immediately after a successful login, with no commands executed. We conjecture that these scanners must save these credentials to be used by another attacking host at a later time. By augmenting the authentication configurations that Cowrie provides and allowing successful credentials to remain ``active" for a short period of time, we found a large number of attackers that had successful first login attempts without any failed ones. We aim to cluster these attackers that we suspect are logging and distributing successful credentials they find during their brute-force scans.

We note that even though we remove commonly tried credentials, there are a few obstacles to our approach. Firstly, there may be an overlap with IPs using the same credential list. Secondly, we have no information to discern which previous IP was responsible for sharing a credential, and it is also possible for multiple IPs to have shared with multiple attackers if a centralized sharing setup exists. Hence, we end up with a many-to-many relationship between these IPs, which we can capture as a graph by drawing edges from each IP successful in their first attempt to all previous IPs that ``discovered'' the credential during a random attempt. Visualizing this data, we find that for all regions, this created a single large connected subgraph, with a few smaller strongly connected sets. Figure \ref{Fig:sharedcreds} shows the graph for the London region.

It is also possible for unrelated IPs to have discovered the same credential. However, we can overcome these false-positive edges by looking for strong relationships between other IPs in the group. Hence, we turned to using a community detection algorithm for clustering, which extracts well connected IPs while dropping spurious edges. Specifically, we used a greedy modularity maximization algorithm \cite{clauset04}, which begins with each node in its own community, and then iteratively joins pairs of communities which increase the modularity of the graph the most. Using this approach, we end up with 12 clusters from the Singapore data, 13 from London and 28 from N. Virginia.

\hypertarget{hassh-fingerprinting}{%
\subsubsection{Hassh Fingerprinting}\label{hassh-fingerprinting}}

Cowrie applies \emph{hassh fingerprinting} \cite{ghiette19} to each SSH session
it handles. Similar to JA3 hashes \cite{althouse19}, a \emph{hassh value} is
generated from the KEY\_MSG\_KEXINIT packets exchanged by the client and
server as part of setting up an SSH channel. The presence and ordering
of the algorithms in the key exchange method, encryption, message
authentication, and compression fields offered by the communicating
machines is concatenated into a string which is MD5 hashed into the
final hassh fingerprint.

According to \cite{ghiette19}, hassh fingerprints are unique enough to provide an accurate estimate of the tool or library used to connect to an SSH server. Thus, we examined it for our problem, as it seemed to be a promising solution to find attackers that use the same SSH tooling. However, we found that this was not enough to confidently establish coordination among numerous attacking IPs. Our data across all regions showed that over 85\% of connections generated one of two predominant hasshes, which both resolved to versions of the highly popular OpenSSH software. Examining the behavior of attacking IPs that generated identical hasshes, we observed different activity once they logged into the honeypot. It makes sense that attackers are content to use the default SSH utility installed on compromised machines to create connections, which end up resolving to the predominant versions of the SSH library installed in the wild.

% In theory, the hassh standard may be most effective from a defensive standpoint, when it can be used to enforce a narrow whitelist of applications in a tightly constrained network environment. 
In theory, the hassh standard may be more effective if there is higher variation in SSH standards and proliferation of different versions of SSH libraries. We found hassh fingerprints to be overly broad for our problem, and not useful to correlate attacker IPs. Hence, we did not use these fingerprints as another independent clustering feature, but we note that improved fingerprints that can identify tooling better can make such methods feasible for clustering attackers in the future.

\hypertarget{consensus-clustering}{%
\subsection{Consensus Clustering}\label{consensus-clustering}}

\hypertarget{tbl:ami-table}{}
\begin{table*}
    \caption{Adjusted mutual information score between each pair of clustering methods for the N. Virginia region}
    \label{tbl:ami-table}
    \centering
    \begin{tabular}{|l|r|r|r|r|r|r||r|}
    \hline
         & Heuristic & Outbound Request & Intervals & Session Activity & Credential List & Shared Credentials & Average AMI \\ 
     \hline\hline
        Heuristic & 1 & 0.483 & 0.558 & 0.605 & 0.782 & 0.742 & 0.634 \\ 
        Outbound Request & 0.483 & 1 & 0.267 & 0.341 & 0.460 & 0.205 & 0.351 \\ 
        Intervals & 0.558 & 0.267 & 1 & 0.380 & 0.534 & 0.405 & 0.429 \\ 
        Session Activity & 0.605 & 0.341 & 0.380 & 1 & 0.563 & 0.448 & 0.467 \\ 
        Credential List & 0.782 & 0.460 & 0.534 & 0.563 & 1 & 0.764 & 0.620 \\ 
        Shared Credentials & 0.742 & 0.205 & 0.405 & 0.448 & 0.764 & 1 & 0.513 \\ 
    \hline
    \end{tabular}
\end{table*}

% \hypertarget{tbl:ami-table}{}
% \begin{table}
% 	\caption{Adjusted mutual information score between each pair of clustering methods for the N. Virginia region\\(HG=Heuristic Grouping, OR=Outbound Request, PI=Probe Interval, SA=Session Activity, CL=Credential List, SC=Shared Credentials, AA=Average AMI)}
% 	\label{tbl:ami-table}
% 	\centering
% 	\begin{tabular}{|l|r|r|r|r|r|r|r|}
% 		\hline
% 		& HG & OR & PI & SA & CL & SC & AA \\ 
% 		\hline\hline
% 		HG & 1 & 0.483 & 0.558 & 0.604 & 0.781 & 0.742 & 0.633 \\ 
% 		OR & 0.483 & 1 & 0.267 & 0.341 & 0.460 & 0.205 & 0.351 \\ 
% 		PI & 0.558 & 0.267 & 1 & 0.380 & 0.533 & 0.405 & 0.428 \\ 
% 		SA & 0.604 & 0.341 & 0.380 & 1 & 0.563 & 0.448 & 0.467 \\ 
% 		CL & 0.781 & 0.460 & 0.533 & 0.563 & 1 & 0.763 & 0.620 \\ 
% 		SC & 0.742 & 0.205 & 0.405 & 0.448 & 0.763 & 1 & 0.512 \\ 
% 		\hline
% 	\end{tabular}
% \end{table}

For our final step, we seek to combine our results from the independent clustering methods to uncover correlations in our attacker activity. As we have seen so far, each cluster set is distinct, and due to lack of ground truth, we have no mechanism to determine the performance of each individual clustering metric. Additionally, our cluster results likely contain differing amounts of overlaps and inaccuracies, as each individual method relies on partial knowledge of the data, as well as our reliance on various assumptions about Internet scanner behavior.

We use the ideas from consensus clustering \cite{monti03, strehl02} to combine our results and overcome errors in the individual clustering methods. Originally devised as an approach to overcome issues in any one clustering of datapoints (e.g., random initializations in clustering algorithms, different parameter settings), consensus clustering has also been shown to overcome various types of noise. Moreover, consensus clustering is a useful, algorithmic method for combining different views of the same data.

As we have seen in previous sections, there are many ways to analyze and cluster IP addresses in honeypot logs. Additionally, different honeypot logs will likely be amenable to other forms of analysis and clustering. We believe consensus clustering is a useful, general technique suitable for such analysis, allowing a straightforward method for fusing information while simultaneously filtering out certain types of noise. While other methods of combining the different clusterings are possible, we focus on consensus clustering for the rest of this paper as an illustrative case study for how one can use this technique to discover coordinated activity in honeypot logs.

We opted to adapt the greedy optimization algorithm \cite{strehl02}, in which all clustering methods are compared to each other by calculating the adjusted mutual information (AMI) between them as well as taking an average across them all. This is followed by an optimization step which assigns each datapoint to every cluster in the starting set in turn, keeping the change only if it increases the average AMI. This effectively expands and combines clusters based on collective agreement, which serves in dropping the errors of the other clustering approaches. If any datapoint changes its label during a sweep of the dataset, it triggers a re-sweep until no changes occur. 

We modify this method for our purposes in two ways -- (1) we allow new clusters to be created, and (2) we fix our heuristic grouping clusters as our starting set for each sweep, as it is our highest confidence result. Table \ref{tbl:ami-table} shows the initial AMI table for the N. Virginia data. Note that AMI ranges between [-1,1], where -1 implies a perfect mismatch, 0 is random, and 1 a perfect match. All of our methods have a positive correlation to each other, showing that they all share similar labels for many IPs, but are also not perfect matches and thus some information that can be learned from each.

% Due to our strict handpicked rules, we consider the result from our heuristic grouping approach as the most reliable, and 
% Since we have a reliable clustering set, the heuristic grouping, The greedy optimization approach also
% allows us to avoid traditional feature-based clustering, where we would
% have to determine an arbitrary way to combine a weighted distance
% between each of our IPs based on how they clustered (if at all) across
% all of our different methods. Instead, by starting from a fairly
% confident set of clusters, we can ask of each IP whether it fits better
% in another existing or new cluster, and only keep the change if enough
% of all the other clustering methods agree on the new assignment.

Running the algorithm over each region, we sweep our London and Singapore clustering results 5 times, with 35 and 22 IPs changing their cluster membership respectively. The N. Virginia data requires 6 sweeps with 69 IPs updating their cluster labels and increases the Average AMI for the Heuristic method shown in the table from .63 to .69. Now with access to additional information, we discover that several small clusters combine into new clusters, and a few IP addresses jump into larger clusters. We end up with 12 clusters from the London data, 16 from the Singapore and 20 from the N. Virginia region, which we note is a significant reduction from our original results from independent heuristic grouping.

\hypertarget{final-result-analysis}{%
\subsection{Final result analysis}\label{final-result-analysis}}

\hypertarget{tbl:result-table}{}
\begin{table}
	\caption{Breakdown of manual analysis}
	\label{tbl:result-table}
	\centering
	\begin{tabular}{|l|r|r|r|r|r|r|r|r|}
		\hline
		Credential List    & \checkmark  & \checkmark & \checkmark & \checkmark &  \checkmark &  &  & \\
		Session Activity &    & \checkmark &  &  & \checkmark &  & \checkmark & \checkmark   \\
		Outbound Request & \checkmark  & \checkmark & \checkmark &  & \checkmark & \checkmark & \checkmark & \checkmark  \\
		Probe Interval   &    &   & \checkmark &   & \checkmark &  & \checkmark &   \\
		\hline
		Clusters Matched    & 16 & 9 & 7 & 5 & 4 & 3 & 2 & 1 \\
		\hline
	\end{tabular}
\end{table}

We analyzed our 48 final clusters by manually evaluating the individual features of each cluster. This process informed us if our consensus clustering was effective in combining information from our independent clustering to form the final result. We did not attempt to combine the clusters across regions, but do note that there were overlapping IPs observed in multiple clusters. We examined each cluster manually by outputting the credential set, session activity, outbound requests and probe intervals of each IP in the cluster and making a determination on which attributes were matched. This exercise not only helped us evaluate the correctness of the clusters based on manual judgement, but also determine the relationship between our chosen features.

% Since we have no ground truth, we rely on manual evaluation that
% leans on domain knowledge to determine if the clusters made sense. We
% explored several resulting clusters and attempted to discern
% the reasoning for their clustering, and whether they matched on
% multiple criteria. Our algorithm built 48 clusters from the data across
% all the regions.  For our evaluation, we manually examined the the credentials, sessions, outbound requests, and probe intervals of each IP in the cluster and made a determination on which criteria they match.

From our examination, we found that 39 of our 48 clusters matched on multiple criteria, 8 matched on a single feature, and one cluster did not match on any. Table \ref{tbl:result-table} shows the breakdown of the 47 clusters, with each column indicating the number of the clusters that matched a certain combination of features. Notice in the table that 36 of our clusters matched on both the credential list as well as outbound request, which makes sense as automated bots will generally attempt to gain access using a pre-configured set of credentials and then execute their proxy request or DDoS attack as soon as they gain access. Out of these, 13 clusters matched on the session activity as well, and we observed that these bots executed the same script on the honeypot in addition to the forwarding request. Another 13 clusters matched on the interval, which shows signs of some scheduling put in place by the attackers controlling the clusters. Additionally, our examination of the various combinations revealed clusters of IPs that tried different credentials but ultimately attempted the same script and outbound requests, as well as ones that executed the same script but finished with different proxy requests. Surfacing such variations in activity from these clusters could provide important information to security teams to help defend their networks.

To further substantiate our result, we also investigated a few interesting clusters. In the first one, we observed the credentials attempted by these IPs were targeting honeypots in particular (e.g., honey:pot, admin:honeypot), and then immediately following with an outbound request once they logged in. In fact, all the IPs in this cluster belonged to two /24 networks owned by the same hosting provider, which makes it feasible to conclude that this activity belonged to singular actor attempting to abuse honeypots. This evidence provides good reason for honeypot users to ensure proper maintenance and configuration of their systems, as they are exploited by capable attackers.

The second set of clusters we investigated were distributed across all three regions, with a match on credential lists and session activity. These were IPs attempting to find Microtik routers (Section \ref{what-do-attackers-do-once-connected}), and we cluster all 1.3K IPs with this activity. Interestingly, most of these IPs only appeared once, and there were just 95 duplicates across the 3 regions. Our final cluster was one of the most consistent sets of activity we saw across our honeypots, which we identified as a singular organization who we refer to as Organization1. These IPs attempted the same set of credentials, session activity as well as outbound request. We found nothing about malicious activity from this organization publicly available, and thus decided to investigate their operations in a case study.

\hypertarget{case-study-organization1}{
\subsubsection{Case Study: Organization1}\label{case-study-organization1}}

Inspecting their primary domain, Organization1 claims to specialize in SEO automation. Despite presenting an outward appearance of nominal legitimacy in their web presence, the solutions that Organization1 offers appear to be targeted directly towards black hat customers. We discovered a large following on black hat forums where users discuss using Organization1 software in illicit ways. As part of our measurement, we noticed several attempts to tunnel connections to destinations controlled by this organization, which return identical pages with the IP address of the requesting machine as the only content. Further examination showed these destinations to be components of a proxy checking software advertised on Organization1's website.

Organization1's main product is a browser automation software that is able to record manual user actions in a browser for subsequent automation. It includes many features aimed at bypassing anti-bot logic including captcha recognition, human profile generation, and mouse/test insert emulation. Another feature of this software is the ability to test and store thousands of proxies for bots to use when performing tasks. We suspect that a significant number of Organization1 IPs are attempting to tunnel traffic through compromised machines to supplement black hat activities using their automation software.

Organization1 operates an official marketplace where clients may buy and sell user-made bots for their software. More buying and selling also occurs on their official forum, where posters can request bots for activities including secretarial work, mass website creation, and mass e-mail account creation. As of this writing, there has been little public notoriety surrounding Organization1 and its software has not gained attention in any cybersecurity articles, blog posts, or academic publications. However, we were able to cluster their bots using our method, using the information gained once they logged in to our honeypot virtual environments. We disclosed all information about Organization1 to our IT security team to disseminate throughout the industry.

% Some examples are a scraper that collects women's profiles on dating websites, an Amazon auto-purchaser/product reviewer, and a Facebook auto-poster.

\hypertarget{discussion}{%
\subsection{Discussion}\label{discussion}}

Due to our collection methodology, i.e., honeypots in various AWS regions with instance IPs, the data we gathered was mostly composed of automated attackers instructed to scan the Internet and capitalize on misconfigured open services. As we observed in our measurement, these scanners had predefined objectives to set up email and web proxies, identify devices, and install cryptominers and scanning software. Because of this, our outbound requests, captured session activity, and credential lists emerged as our strongest features to cluster these attackers. Honeypot data collection on different networks may capture more sophisticated and/or human activity where our more advanced techniques for matching session activity based on command capabilities, probe intervals, and shared credentials could expose more hidden similarities between attackers.

We also consider the existence of NAT devices and IP churn in our data and its effects on our clustering. In the case where multiple devices utilize the same IP for the same activity, this does not affect any of our independent clusters, except for extra connections that may impact our probe interval calculations. However, if multiple devices elicit different types of behavior on our honeypots from a single IP, this can affect our data representations. In addition to some effect on probe intervals, these IPs would add extra edges to the graph for outbound requests, create larger sets of attempted credentials, and different activity in the sessions would cause multiple mappings for the related IPs. This would generally result in a lower similarity score for these IPs, and our clustering algorithms would discard them from the clusters unless one type of activity was heavily favored. Alternatively, one could automatically identify nodes that belong to extremely different clusters across different types of activity by examining the impact of individual nodes on normalized mutual information via a form of multi-view anomaly detection \cite{liu12}. We leave such experiments for future work.

Finally, our consensus clustering technique allows us to have a flexible mechanism to combine results from many different approaches. We can avoid defining arbitrary weights to combine a list of features that may not always be present in different slices of the data, and allows clusters to grow and combine if enough of the participating clustering techniques agree on a re-label for a datapoint. This algorithm also allows easy addition of new independent features, which may be speculative (e.g., the shared credentials), but could be the catalyst needed to connect two disparate clusters. Thus, our technique can be expanded upon to suit the types of networks, protocols, and honeypot capabilities studied in future measurements.

\hypertarget{ethical-considerations}{
\section{Ethical Considerations}\label{ethical-considerations}}

To setup and complete this measurement, we followed established good practices for research on networking systems. As in prior work \cite{barron17,zeng14}, we avoided the use of our institution's infrastructure and performed our measurements on a third party platform (AWS), which allows the installation of passive honeypots as long as they are not abused to attack other infrastructure. Hence, we disabled all outbound connections from our honeypot machines as well as other features that could cause collateral damage (e.g., comment boxes on honey webpages). While the ethics of engaging with honeypots and attackers for networking research has been discussed in the past \cite{barron17, vetterl19}, we avoided all engagement and maintained a completely passive presence throughout our measurement.

\hypertarget{conclusion}{%
\section{Conclusion}\label{conclusion-and-future-work}}

In this work, we investigated the use of multiple honeypots for characterizing network attacks on the open Internet. We measured various statistics about the attacks we observed and what the attackers intended to achieve, and used this information to identify features we could use to cluster them. Focusing on data gathered on our SSH medium-interaction honeypots, we developed a novel approach that combines the results of multiple clustering algorithms to uncover correlations between attacker activity. Our results showed that our attackers can indeed be clustered using a combination of multiple features, and our technique can be used in future network measurements and security systems to help identify attacking IPs that may be controlled by a single operator.

\bibliographystyle{IEEEtranS}
\bibliography{IEEEabrv,references}

\clearpage

\appendices

\section{Examples of session activity on Cowrie}\label{session-activity-appendix}

Activity from Mirai bots on Cowrie honeypots.

\vspace{5mm}
\begin{minipage}{\textwidth}
\begin{Verbatim}[frame=lines,fontsize=\scriptsize,framesep=3mm]
$ /gisdfoewrsfdf	
$ /bin/busybox cp; /gisdfoewrsfdf	
$ mount ;/gisdfoewrsfdf	
$ echo -e '\x47\x72\x6f\x70/' > //.nippon; cat //.nippon; rm -f //.nippon	
$ echo -e '\x47\x72\x6f\x70/tmp' > /tmp/.nippon; cat /tmp/.nippon; rm -f /tmp/.nippon	
$ echo -e '\x47\x72\x6f\x70/var/tmp' > /var/tmp/.nippon; cat /var/tmp/.nippon; rm -f /var/tmp/.nippon	
$ echo -e '\x47\x72\x6f\x70/' > //.nippon; cat //.nippon; rm -f //.nippon	
$ echo -e '\x47\x72\x6f\x70/lib/init/rw' > /lib/init/rw/.nippon; cat /lib/init/rw/.nippon; rm -f /lib/init/rw/.nippon	
$ echo -e '\x47\x72\x6f\x70/proc' > /proc/.nippon; cat /proc/.nippon; rm -f /proc/.nippon	
$ echo -e '\x47\x72\x6f\x70/sys' > /sys/.nippon; cat /sys/.nippon; rm -f /sys/.nippon	
$ echo -e '\x47\x72\x6f\x70/dev' > /dev/.nippon; cat /dev/.nippon; rm -f /dev/.nippon	
$ echo -e '\x47\x72\x6f\x70/dev/shm' > /dev/shm/.nippon; cat /dev/shm/.nippon; rm -f /dev/shm/.nippon	
$ echo -e '\x47\x72\x6f\x70/dev/pts' > /dev/pts/.nippon; cat /dev/pts/.nippon; rm -f /dev/pts/.nippon	
$ /gisdfoewrsfdf	
$ cat /bin/echo ;/gisdfoewrsfdf
\end{Verbatim}
\end{minipage}\hfill

\vspace{5mm}
Other common command sequences:

\vspace{5mm}
\begin{minipage}{\textwidth}
\begin{Verbatim}[frame=lines,fontsize=\scriptsize,label=Microtik,framesep=3mm]
$ /ip cloud print,
$ ifconfig,
$ uname -a,
$ cat /proc/cpuinfo,
$ "ps | grep [Mm]iner",
$ "ps -ef | grep [Mm]iner",
$ ls -la /dev/ttyGSM* /dev/ttyUSB-mod* /var/spool/sms/* /var/log/smsd.log /etc/smsd.conf* /usr/bin/qmuxd 
> /var/qmux_connect_socket /etc/config/simman /dev/modem* /var/config/sms/*,
$ echo Hi | cat -n
\end{Verbatim}

\begin{Verbatim}[frame=lines,fontsize=\scriptsize,framesep=3mm]
$ echo "cd /tmp; rm -f *.sh; wget http://46.246.41.29/wget.sh || curl http://46.246.41.29/curl.sh -o curl.sh; 
> chmod +x *.sh; ./wget.sh; ./curl.sh" | sh
$ cd /tmp; rm -f *.sh; wget http://46.246.41.29/wget.sh || curl http://46.246.41.29/curl.sh -o curl.sh; 
> chmod +x *.sh; ./wget.sh; ./curl.sh\n
\end{Verbatim}

\begin{Verbatim}[frame=lines,fontsize=\scriptsize,framesep=3mm]
$ unset HISTORY HISTFILE HISTSAVE HISTZONE HISTORY HISTLOG WATCH ; history -n ; export HISTFILE=/dev/null ; 
> export HISTSIZE=0; export HISTFILESIZE=0 ; rm -rf /var/log/wtmp ; rm -rf /var/log/lastlog ; rm -rf /var/log/secure; 
> rm -rf /var/log/xferlog ; rm -rf /var/log/messages ; rm -rf /var/run/utmp ; touch /var/run/utmp ; 
> touch /var/log/messages ; touch /var/log/wtmp ; touch /var/log/messages ; touch /var/log/xferlog ; 
> touch /var/log/secure ;  touch /var/log/lastlog ; rm -rf /var/log/maillog ; touch /var/log/maillog ; 
> rm -rf /root/.bash_history ; touch /root/.bash_history ; history -r ,
$ uname,
$ free -m,
$ ps -x,
$ cat /proc/cpuinfo
\end{Verbatim}

\begin{Verbatim}[frame=lines,fontsize=\scriptsize,framesep=3mm]
$ shell,
$ uname -r,
$ id,
$ id,
$ ls -la /usr/bin/curl,
$ ps ax|grep dhc,
$ uname -s -m,
$ cat /proc/version;cat /proc/cpuinfo
\end{Verbatim}

\end{minipage}\hfill

\clearpage

\section{Ransomware message on PSQL Honeypot}\label{psql-activity-appendix}
\begin{Verbatim}[frame=lines,fontsize=\scriptsize,framesep=3mm]
Hello, 
I am a security researcher from Sweden, having interest on web security and other focus areas. 
Your database was removed by a 3rd party and files were backed up to their cloud hosting storage. 
It is scheduled to be sold online. I accidently discovered this dedicated 
cloud storage and was able to secure the files. 

To retrieve the original dump file and prevent public leaking of the database: 
Please send exactly 0.1 Bitcoin (BTC) to the following Bitcoin address: 17UT9ESnwpmfMbGr4oXFqoQfSn8Bn]XqFg 

Please contact by email one hour after payment complete: 7b2b7e74c922@mailinator.com 
and let me know the email or ssh account to upload the file. 
I will upload the original dump file The.sql.gz with separate dump files per table
(full structure dump including schema, views, triggers and data section). 
I will also shred remove any files and terminate the cloud hosting account. 
You can email to verify the data dump: 7b2b7e74c922@mailinator.com. 
Please make sure to include your server ip, database name and table name(s). 
I remove original dump in 24hrs after payment if recovery successful, 
you will have to import your data with database cli tools. 

To buy bitcoin instantly you can use paxful.com like services. 
* Should you don"t need the data, i will leave the files as it is, 
your database will be sold and(or) exposed online or used otherwise. 
\end{Verbatim}

\section{Examples of activity on miniprint honeypot}\label{miniprint-activity-appendix}

\footnotesize{
\hypertarget{tbl:miniprint-print-activity}{}
% \begin{table*}
%     \caption{Example print activity on miniprint honeypot}
%     \label{tbl:miniprint-print-activity}
    \centering
    \begin{tabular}{l|p{13cm}|p{1.7cm}}
    \hline
        Date & Printed Message & Activity \\ 
    \hline\hline
        Sep-2019 & \texttt{You have been hacked!You have been hacked!You have been hacked!You have   been hacked!} (54 times) & Vandalism \\ 
        \hline
        Oct-2019 & \texttt{(587) 779-8484 please text this number it is HP support, and tell them   to stop making their printers so hackable, with love from prisma team <3} & Vandalism \\ 
        \hline
        Dec-2019 & \texttt{FREEHKSAVEKOREA} (printed multiple times over two weeks) & Hacktivism \\ 
        \hline
        Jun-2020 & \texttt{\#OPJUSTICE4FLOYD} (printed multiple times during the month) & Hacktivism \\ 
        \hline
        Jun-2020 & \texttt{HI THERE$\backslash$n$\backslash$nYOU$\backslash$xe2$\backslash$x80$\backslash$x99RE PROBABLY WONDERING HOW AND WHY YOUR PRINTER PRINTED OUT THIS MESSAGE$\backslash$n$\backslash$nYOUR PRINTER HAS THE PORT 9100 RIGHT OPEN   AND I MANAGED TO CREATE AN AUTOMATED SCRIPT TO SEND THIS MESSAGE TO PRINTERS WITH THE SAME VULNERABILITY ALL OVER THE WORLD$\backslash$n$\backslash$nI RECOMMEND YOU TO RESEARCH HOW TO PREVENT LEAVING YOUR PRINTERS VULNERABLE LIKE THIS$\backslash$n$\backslash$nTHIS MESSAGE IS A COURTESY OF SP4DE$\backslash$n$\backslash$nTWEET A PICTURE TO \#PRINTERHACKEDBYSP4DE$\backslash$n$\backslash$n sp4dexzyz@protonmail.com} & Vandalism and Advertisement \\ 
        \hline
        Sep-2020 & \texttt{NetSystemsResearch studies the availability of various services across   the internet. Our website is netsystemsresearch.com} & Advertisement \\ 
    \hline
    \end{tabular}
% \end{table*}
}

\section{List of top 100 common credentials on Cowrie}

\begin{minipage}{\textwidth}
\begin{Verbatim}[frame=lines,framesep=3mm]
admin:                        root:default         root:nosoup4u       bananapi:bananapi  
admin:admin                   root:000000          admin:motorola      xbian:raspberry    
admin:admin123                root:welc0me         root:anko           root:abcd1234      
admin:123456                  root:123             root:toor           root:1q2w3e4r      
pi:raspberry                  root:ubnt            admin:changeme      vyos:vyos          
pi:raspberryraspberry993311   root:0000            root:12345678       root:1234567       
admin:password                root:uClinux         admin:pfsense       root:qazwsxedc     
root:admin                    admin:admin1234      root:admintrup      root:test          
support:support               root:Zte521          root:!@             1234:1234         
root:root                     root:system          root:xmhdipc        support:support123 
root:123456                   root:dreambox        guest:123456        root:abc123        
user:user                     default:default      test:123456         root:1qaz2wsx      
ubnt:ubnt                     root:openelec        admin:manager       pi:bananapi        
root:1234                     postgres:postgres    user:1              testuser:testuser  
root:password                 root:1               usuario:usuario     hadoop:123123      
root:12345                    guest:guest          admin:123           pi:bananapi        
test:test                     root:raspberrypi     ftpuser:ftpuser     testuser:testuser  
admin:1234                    osmc:osmc            root:root123        hadoop:123123      
admin:12345                   admin:1111           vyatta:vyatta       root:p@ssw0rd      
root:111111                   admin:7ujMko0admin   ftp:ftp             postgres:123456    
admin:&                       oracle:oracle        debian:temppwd      backup:backup      
user:1234                     admin:aerohive       root:123456789      git:git            
ubuntu:ubuntu                 root:waldo           admin:admin1        ftp:123456         
admin:default                 root:seiko2005       operator:operator   username:password  
service:service               root:rpitc           root:P@ssw0rd       root:root1234      
\end{Verbatim}
\end{minipage}\hfill

\end{document}